\documentclass{ws-procs975x65}
\usepackage{graphicx}
\usepackage{amsmath}
\usepackage{amssymb}
\usepackage{enumerate}
\usepackage{epsfig}
\usepackage{verbatim}
\usepackage{array}

\makeatletter

\def\vereq#1#2{\lower3pt\vbox{\baselineskip0.5pt\lineskip0.5pt
\ialign{$\m@th#1\hfill##\hfil$\crcr#2\crcr\sim\crcr}}}

\makeatother

\newcommand{\regularwidthofpicture}{0.7\textwidth}

\begin{document}

\title{Anisotropies in Ultrahigh Energy Cosmic Rays\footnote{
  \uppercase{I}nvited talk at
  the 10th \uppercase{M}arcel \uppercase{G}rossmann \uppercase{M}eeting
  in \uppercase{R}io de \uppercase{J}aneiro,
  20--26 \uppercase{J}uly 2003}}

\author{John Swain}
\address{
Department of Physics,
Northeastern University, Boston, MA 02115\\
e-mail: {\tt john.swain@cern.ch}}


\maketitle

\abstracts{The present status of anisotropy studies for the highest energy 
cosmic rays is presented including the first full sky survey. Directions and prospects for the future are also discussed in light of new statistical methods and the last quantities of data expected in the near future from the Pierre Auger Observatory.}

\section{Introduction}

Ultrahigh energy cosmic rays (UHECR's) are among the most enigmatic phenomena in the 
universe.\cite{Bhattacharjee:1998qc} In the mid-60's Greisen, Zatsepin, 
and Kuzmin (GZK) pointed out that 
ultra high energy protons 
interact with the all-pervading cosmic microwave background via photopion 
production.\cite{Greisen:1966jv} Strictly speaking, protons with energies 
$> 10^{19.6}$~eV 
have a mean interaction length $< 6$~Mpc and an inelasticity 
of about $20\%$ per interaction. Consequently, the popular astronomical picture, namely proton 
``bottom-up'' acceleration in extragalactic objects, predicts a sharp suppression of the cosmic 
ray intensity somewhat beyond $10^{19.8}$~eV. For heavy nuclei, the giant dipole resonance can be 
excited at similar total energies and hence iron nuclei do not survive fragmentation over 
comparable path lengths.\cite{Puget:nz} 

The existence of cosmic rays with energies exceeding $10^{20}$~eV has been 
observed by the Volcano Ranch,\cite{Linsley} the 
Haverah Park,\cite{Ave:2000nd} 
the Sydney University 
Giant Airshower Recorder (SUGAR),\cite{Anchordoqui:2003gm} 
the Akeno Giant Air Shower Array (AGASA),\cite{Takeda:2002at} and 
the Fly's Eye\cite{Bird:1994uy} experiments.  
Because of the GZK cutoff, these cosmic rays should be produced in nearby 
active astronomical objects. Such high energy ``stars'' have been searched for in the arrival direction of these events but no clear candidates were found.

Of course, there are ways to avoid the distance restriction 
imposed by the GZK effect. For instance, there could be a ``top-down'' 
mechanism where 
(charged and/or neutral) supermassive $X$-particles are produced at extreme energies. Sources 
of these exotic particles could be topological defects left 
over from early universe phase 
transitions associated with the spontaneous symmetry breaking that underlies unified models of high energy 
interactions,\cite{Bhattacharjee:1991zm} or else some long-lived metastable super-heavy relic particles 
produced through vacuum 
fluctuations during the inflationary stage of the 
universe.\cite{Berezinsky:1997hy} From time to time, 
the energy stored in a single $X$ can be released in the form of massive quanta that typically produce 
jets of hadrons well above the highest observed energies. However, it is noteworthy that there can be a 
problem with some top-down 
interpretations. Specifically, the $X$-particle cascades may produce a rather large flux of energetic 
photons and neutrinos, possibly in excess of the upper limits already established.\cite{Protheroe:1996pd}
The lack of plausible nearby astrophysical sources has also 
encouraged the idea of positing undiscovered neutral hadrons,\cite{Farrar:1996rg} as well 
as mechanisms which are able to break the GZK barrier.\cite{Fargion:1997ft} Even though sufficiently 
heavy particles would avoid photopion production (the threshold energy varies as the square of the mass 
of the first resonant state), the existence of these particles 
now appears to be excluded by laboratory experiments.\cite{Albuquerque:1994xi} The only standard model 
particle that can reach our galaxy from high redshift sources without significance loss of energy is the 
neutrino. However, the expected event rate for early development of a neutrino shower is less than that 
of an electromagnetic or hadronic interaction by several orders of magnitude, even if ones takes into 
account non-standard graviton mediated interactions.\cite{Kachelriess:2000cb}

The distribution of arrival directions is perhaps the most helpful observable in yielding clues 
about cosmic ray origin. This may come either from clustering on a small angular scale 
that identifies discrete sources,\cite{Hayashida:bc} or else as a large-scale celestial pattern that 
characterizes a particular class of potential sources.\cite{Berezinsky:1998rp} Along these lines, data 
observed by the AGASA,\cite{Hayashida:1998qb} the SUGAR,\cite{Bellido:2000tr} and the Fly's Eye~\cite{Bird:1998nu} experiments 
show an excess flux of cosmic rays from a direction near the Galactic center up to about $10^{18.5}$~eV, 
but there is evidence for Galactic plane avoidance above that energy. Such an effect can be easily explained 
if cosmic rays are mostly protons and nuclei, because their magnetic rigidity  increases with energy and 
so one expects the angular width of the Galactic plane as seen in protons and nuclei would shrink 
slowly with rising energy. Moreover, the events yielding the observed 
anisotropy are concentrated in a limited energy range. This is very suggestive of neutrons as candidate primaries, 
because the directional signal requires relatively-stable neutral primaries, and time-dilated neutrons can reach 
Earth from typical Galactic distances when the neutron energy exceeds $10^{18}$~eV. Arguably, if the Galactic messengers 
are neutrons, then those with energies below $10^{18}$~eV will decay in flight, providing a flux of cosmic antineutrinos 
above a TeV which is {\it observable} at a kilometer-scale neutrino observatory.\cite{Anchordoqui:2003vc} 
A measurement of this flux can serve to identify the first extraterrestrial point source of TeV antineutrinos.

All in all, the data around $10^{18.5}$~eV suggests that a new population of 
cosmic rays with extragalactic origin begins to dominate the more steeply falling Galactic population. 
In the extragalactic sway, the evidence for anisotropy patterns is suggestive but 
statistically very weak. On the one hand, correlations with the local structure of galaxies have been 
reported,\cite{Stanev:1995my} suggesting that all cosmic rays with energies 
$> 10^{18.5}$~eV are 
emitted by nearby astrophysical sources. On the other hand, possible correlations with high redshift 
astrophysical objects, which most likely indicate new physics, are also under debate.\cite{Farrar:1998we} 
Clearly, a positive identification of ultra high energy cosmic ray sources from the distribution of 
arrival directions requires a careful study of such distribution over the full celestial sphere. 
The latter is the main inspiration for this talk, which is 
organized as follows. In order to set the stage for the full-sky coverage discussion, in Sec. 2 I will   present 
an overview of available statistics and existing anisotropy studies. In particular, I will concentrate  
on the SUGAR and AGASA experiments and discuss in some detail the exposure of these ground arrays.  
Next, in Sec. 3, I will review the main properties of the 
angular power spectrum\cite{Peebles} and study cosmic ray anisotropies from an expansion on spherical harmonics for modes out 
to $\ell =5$.\cite{Anchordoqui:2003bx} In Sec. 4 a numerical likelihood 
approach to the determination of cosmic ray anisotropies is 
presented.\cite{Hojvat:2003su} This 
method  offers many advantages over other approaches: It allows 
a wide range of statistically meaningful hypotheses to be compared even when 
full sky coverage is unavailable, can be readily extended in order to include 
measurement errors, and makes maximum unbiased use of all available 
information. Finally, I will summarize and present the 
conclusions in Section 6.

\section{Experimental data sets}

The SUGAR array was operated from January 1968 to February 1979 in  New South Wales (Australia) at a 
latitude of $30.5^\circ$ South and longitude $149^\circ 38'$ East.\cite{Winn:un} The array consisted 
of 47 independent  
stations on a rectangular grid covering an area $S \approx 70$~km$^2$. The primary energy was determined 
from the total number of muons, $N_\mu$, traversing the detector at the measured zenith angle $\theta$. 
The total aperture for incident zenith angles between $\theta_1$ and $\theta_2$ was found to be
\begin{equation}
A =  \int_{\theta_1}^{\theta_2} S \,\,p(N_\mu, \theta) \, \cos \theta \, d\Omega \,.
\end{equation}
Here, $p(N_\mu, \theta)$ is the probability that a shower falling within the physical area
was detected, $S \cos \theta$ is the projected surface of the array in the shower plane, and 
$d\Omega$ is the acceptance solid angle. The SUGAR Collaboration reports\cite{Winn:un}
a reasonable accuracy in assessing the shower parameters up to $\theta = 73^\circ$. 
The estimated angular uncertainty for showers that triggered 5 or more stations is reported 
as $3^\circ \sec \theta$.\cite{Winn:un} However, the 
majority of events were only viewed by 3 or 4 stations, 
and for these the resolution appears to be as poor as $10^\circ$.\cite{Kewley:zt} Of particular 
interest for this analysis,
\mbox{$p(N_\mu > 10^8, \theta < 55^\circ) \approx 0.85$},\cite{Bell:gp} yielding a total aperture $A
\approx 125$~km$^2$~sr. This provides an exposure reasonably matched to that of AGASA, which is 
described next.

The AGASA experiment occupies farm land near the village of Akeno (Japan) at a longitude 
of $138^\circ 30'$ East and  latitude $35^\circ 30'$ North.\cite{Chiba:1991nf} The array, 
which consists of 111 
surface detectors deployed over an area of about 100 km$^2$, has been running since 1990. 
About 95\% of the surface detectors were operational from March to December 1991, and
the array has been fully operational since then.  A prototype detector operated from 1984 to 1990  
and has been part of AGASA since 1990.\cite{Teshima:1985vs}
The aperture for events with primary zenith angle $0^\circ < \theta < 45^\circ$ and energies 
beyond $10^{19.25}$~eV is found to be $A \approx 125$~km$^2$ sr.\cite{Chiba:1991nf} 
The angular resolution for these events is $1.6^\circ$.\cite{Takeda:1998ps}

\begin{figure}[tbhp]
  \begin{center}
	\leavevmode \epsfig{file=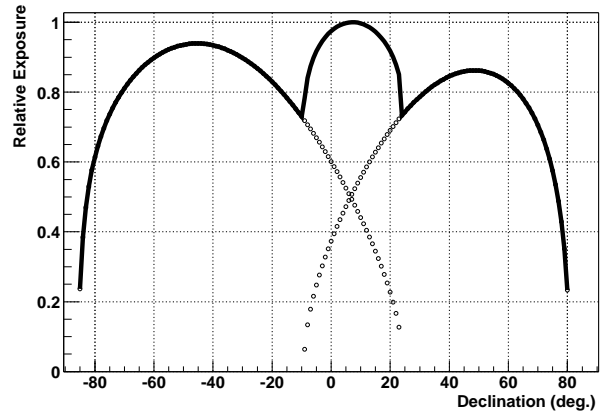,width=\regularwidthofpicture}
    \caption{Declination dependence of SUGAR and AGASA relative apertures (dotted). The solid line indicates  
the combined relative aperture. 
      \label{fig1} }
    \end{center}
  \end{figure}

A detector at latitude $a_0$ that has continuous operation with constant 
exposure in right ascension and is fully efficient for $\theta < \theta_{\rm max}$ has relative exposure 
with the following dependence on declination\cite{Sommers:2000us} 
\begin{equation}
\omega(\delta) \propto (\cos a_0\,\,\cos\delta\,\,\sin\alpha_{\rm max} + \alpha_{\rm max} 
\,\,\sin a_0\, \,\sin \delta)\,\,,
\label{omeguita}
\end{equation}
where $\alpha_{\rm max}$, the local hour angle at which the zenith angle becomes 
equal to $\theta_{\rm max}$, is given by
\begin{equation}
\alpha_{\rm max} = \left\{ \begin{array}{ll}
0 & {\rm if}\,\,\,\xi > 1 \\
\pi & {\rm if} \,\,\, \xi < -1 \\
\cos^{-1}\,\,\xi & {\rm otherwise}
\end{array} \right.
\end{equation}
with
\begin{equation}
\xi \equiv \frac{\cos \theta_{\rm max} - \sin a_0\,\,\sin \delta}{\cos a_0\,\,\,\cos\delta}\,\,.
\end{equation} 
The resulting declination dependence for SUGAR and AGASA together with the combined aperture 
is given in Fig.~\ref{fig1}. 

\begin{figure}[tbhp]
  \begin{center}
	\leavevmode \epsfig{file=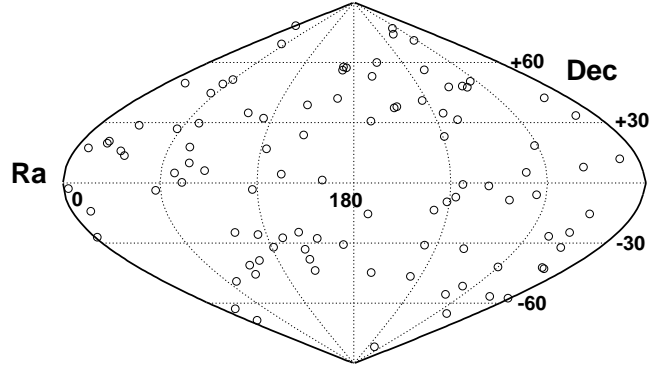,width=\regularwidthofpicture}
    \caption{
Arrival direction of the 99 events observed above $10^{19.6}$~eV by the SUGAR ($\theta < 55^\circ$) 
and the AGASA ($\theta < 45^\circ$) experiments (equatorial coordinates 
B.1950).
      \label{fig2} }
    \end{center}
  \end{figure}

As one can readily see in Fig.~\ref{fig1}, the combined aperture 
of SUGAR and AGASA arrays is nearly uniform over the entire sky. The expected 
event rate is found to be
\begin{eqnarray}
\frac{dN}{dt} &  = &  A\, \int_{E_1}^{E_2}\, E^3 J(E)\, \frac{dE}{E^3} \nonumber \\ 
 & \approx & \frac{A}{2} \,\langle E^3\, J(E) \rangle\, \left[ \frac{1}{E_1^2} - \frac{1}{E_2^2} \right] \,\,,\label{flux}
\end{eqnarray}
where $\langle E^3 J(E) \rangle \approx 10^{24.6}$~eV$^2$ m$^{-2}$ s$^{-1}$ sr$^{-1}$ stands for the 
observed ultra high energy cosmic ray flux, which has a cutoff at $E_2 = 10^{20.5}$~eV.\cite{Bhattacharjee:1998qc} Thus, 
in approximately 10~yr of running each of 
these experiments should collect $\approx 50$ events above $E_1 = 10^{19.6}$~eV, arriving with a zenith 
angle $< \theta_{\rm max}$. Here, $\theta_{\rm max} = 45^\circ$ 
for AGASA and $\theta_{\rm max} = 55^\circ$ for SUGAR. Our sub-sample for the full-sky anisotropy 
search consists of the 50 events detected by AGASA from May 1990 
to May 2000,\cite{Hayashida:2000zr}
and the 49 events detected by SUGAR  with $\theta < 55^\circ$.\cite{Winn:un} Note that we consider the full data sample for the 11 
yr lifetime of SUGAR (in contrast to the 10 yr data sample from AGASA). This roughly compensates for the time variation of 
the sensitive area of the experiment as detectors were deployed or inactivated for 
maintenance. The arrival directions of the 99 events are plotted in Fig.~\ref{fig2} (equatorial coordinates 
B.1950).

\section{Correlations and Power Spectrum}

We begin this section with a general introduction to
the calculation of the angular power spectrum and the determination of
the expected size of intensity fluctuations.  The technique is then
applied to the AGASA and SUGAR data in order to check for
fluctuations beyond those expected from an isotropic distribution.

Let us start by defining the directional phase space of the
angular distribution of cosmic ray events in equatorial coordinates, $(\alpha, \delta)$. 
(i) The direction of the event is described by a unit vector
\begin{equation}
{\bf n}=\sin\delta \,({\bf i}\,\cos\alpha +{\bf j}\, \sin\alpha )+{\bf k}\,\cos\delta \,\,;
\label{CP1}
\end{equation}
(ii) The solid angle is given by
\begin{equation}
d^2{\bf n}=\sin\delta \,\,d\delta \,d\alpha \,\,;
\label{CP2}
\end{equation}
(iii) The delta function for the solid angle is defined as
\begin{equation}
\delta ({\bf n},{\bf n}^\prime ) =
\delta (\cos \delta -\cos \delta^\prime )\,
\sum_{m=-\infty }^\infty
\delta (\alpha -\alpha^\prime +2\pi m)\,,
\end{equation}
so that, as usual,
\begin{equation}
f({\bf n}) = \int
\delta ({\bf n},{\bf n}^\prime )\,\,f({\bf n}^\prime )\,\,d^2{\bf n}^\prime \,\,;
\label{CP3}
\end{equation}
(iv) The probability distribution
\begin{math} P({\bf n})d^2{\bf n} \end{math}
of events can be employed for the purpose of computing the averages
\begin{equation}
\overline{f}=\int f({\bf n})\,\,P({\bf n})\,\,d^2{\bf n}\,\,;
\label{CP4}
\end{equation}
Finally, (v) for a sequence of \begin{math} N \end{math} different
cosmic ray events \begin{math}({\bf n}_1,\ldots ,{\bf n}_N)\end{math}
one may assume an independent distributions for each event, i.e.
\begin{equation}
P_N({\bf n}_1,\ldots ,{\bf n}_N)\prod_i^N d^2{\bf n}_i=
\prod_i^N \{P({\bf n}_i)\,d^2{\bf n}_i\}\,\,.
\label{CP5}
\end{equation}

For a sequence of events
\begin{math}({\bf n}_1,\ldots ,{\bf n}_N)\end{math}
let us describe the angular intensity as
the random variable
\begin{equation}
I({\bf n})=\frac{1}{ N}\sum_{j=1}^N \,  \delta ({\bf n},{\bf n}_j)\,\,.
\label{CP6}
\end{equation}
From Eqs.~(\ref{CP5}) and (\ref{CP6}) it follows that
\begin{eqnarray}
\overline{I({\bf n})} & = &
\int \ldots \int
I({\bf n})\,P_N({\bf n}_1,\ldots ,{\bf n}_N)\,\prod_i^N d^2{\bf n}_i \nonumber \\
&  = & P({\bf n}).
\label{CP7}
\end{eqnarray}
The two point correlation function
\begin{math} G({\bf n},{\bf n}^\prime )
=\overline{I({\bf n})I({\bf n}^\prime )} \end{math}
is defined via
\begin{eqnarray}
G({\bf n},{\bf n}^\prime )&=&\int \ldots \int
I({\bf n})\ I({\bf n}^\prime ) \
P_N({\bf n}_1,\ldots ,{\bf n}_N) \ \prod_i^N d^2{\bf n}_i \nonumber \\
 &=& \frac{1}{N}\,\delta ({\bf n},{\bf n}^\prime )\,P({\bf n})+
\left(1-\frac{1}{N}\right)\,P({\bf n})\,P({\bf n}^\prime ) \,\,. \nonumber\\
 & & 
\label{CP8}
\end{eqnarray}

The ``power spectrum'' of the correlation function is determined by
the eigenvalue equation
\begin{equation}
\int G({\bf n},{\bf n}^\prime )\,\,\psi_\lambda ({\bf n}^\prime )\,\,
d^2{\bf n}^\prime=\lambda \,\,\psi_\lambda ({\bf n}).
\label{CP9}
\end{equation}
In this regard it is useful to introduce Dirac notation to indicate the inner product
\begin{equation}
\left<\psi|\psi\right>=
\int \psi^*({\bf n})\,\,\psi({\bf n})\,\,d^2{\bf n}\,\,.
\label{CP10}
\end{equation}
With this in mind, Eq.~(\ref{CP9}) reads
\begin{equation}
G \,\left|\psi_\lambda \right>=\lambda \,\left|\psi_\lambda \right>.
\label{CP11}
\end{equation}

In the limit of a large number of events
\begin{math} N\to \infty  \end{math},
\begin{equation}
\lim_{N\to \infty}G({\bf n},{\bf n}^\prime )
\equiv G_\infty ({\bf n},{\bf n}^\prime )
=P({\bf n})P({\bf n}^\prime )\,,
\end{equation}
or equivalently,
\begin{equation}
\hat{G}_\infty = \left|P\right>\left<P\right| \,.
\label{CP12}
\end{equation}
In such a limit, fluctuations can be neglected and we
find only {\em two
possible values} in the spectrum: (i) There is a
non-degenerate non-zero
eigenvalue
\begin{equation}
\hat{G}_\infty \left|P\right> = \lambda_\infty
\left|P\right>\,,
\end{equation}
with
\begin{equation}
\lambda_\infty = \left<P|P\right> = \int P^2({\bf
n})d^2 {\bf n}.
\label{CP13}
\end{equation}
(ii) For every state \begin{math} \left| f \right>
\end{math}
orthogonal to \begin{math} \left| P \right>
\end{math}
with mean value \begin{math}
\bar{f}=\left<P|f\right>=0 \end{math},
there exists a {\em zero eigenvalue} in the power
spectrum
\begin{equation}
\hat{G}_\infty \left|f\right> = 
\left|P\right>\left<P|f\right>
=\left\{\int P f d^2{\bf n}\right\}\left|P\right>
=\bar{f}\left|P\right> = 0 \,\,.
\label{CP14}
\end{equation}
Let us  now turn  to consider the effects of finite \begin{math}
N \end{math}. Defining the fluctuations in the intensity by
\begin{equation}
\Delta I({\bf n})=I({\bf n})-\overline{I({\bf n})}
=I({\bf n})-P({\bf n}),
\label{CP15}
\end{equation}
the two point correlation function can be re-written as
\begin{eqnarray}
G({\bf n},{\bf n}^\prime ) &  = &
\overline{I({\bf n})I({\bf n}^\prime )} =
\overline{I({\bf n})}\ \overline{I({\bf n}^\prime )}
+\overline{\Delta I({\bf n})\Delta I({\bf n}^\prime )}
\nonumber \\
 & = & G_\infty ({\bf n},{\bf
n}^\prime )
+\overline{\Delta I({\bf n})\Delta I({\bf n}^\prime )}\,,
\end{eqnarray}
with
\begin{equation}
\overline{\Delta I({\bf n})\Delta I({\bf n}^\prime )}
 = \frac{1}{N}\,\left[\, \delta ({\bf n},{\bf n}^\prime
)P({\bf n})
-P({\bf n})P({\bf n}^\prime )\,\right] \,,
\label{CP16}
\end{equation}
where Eq.~(\ref{CP8}) has been invoked. Putting all this together, some general results 
follow: (i) For the \mbox{$N\to\infty$} case, there is
only one state
with a finite eigenvalue \begin{math} \lambda_\infty
\end{math}, while
the rest of the power spectrum corresponds to
\begin{math} \lambda =0  \end{math}. (ii) For finite
\begin{math} N \end{math}, Eq.~(\ref{CP16}) implies that
the fluctuations are of order
\begin{math} N^{-1} \end{math}. The power
spectrum for large \begin{math} N \end{math} then has
one eigenvalue of order unity and the rest of the eigenvalues are of
order
\begin{math} N^{-1} \end{math}.

\begin{figure}[thpb]
\begin{center}
\includegraphics[width=0.7\textwidth]{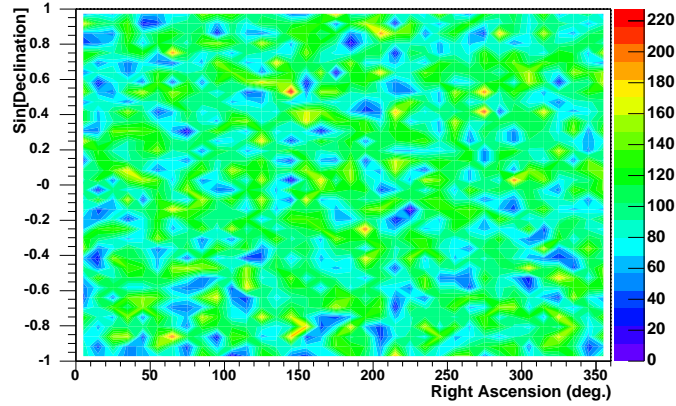}
\caption{Intensity of the cosmic ray sky in equatorial coordinates as seen by the SUGAR and AGASA arrays.}
\label{fig:intensity}
\end{center}
\end{figure}

Now, for an isotropic distribution of ${\bf n}$,
\begin{equation}
\widetilde{P} ({\bf n}) = \frac{1}{4\pi } \,\,,
\end{equation}
and the two point correlation function reads,
\begin{equation}
\widetilde{G} ({\bf n},{\bf n}^\prime ) =
\frac{1}{4\pi N}\,\,\delta ({\bf n},{\bf n}^\prime )+
\frac{1}{(4\pi )^2}\,\left({1-\frac{1}{N}}\right) \,\,.
\label{CP17}
\end{equation}
The eigenvalue problem is solved by employing spherical
harmonics~\cite{ylm} 
\begin{equation}
\int \widetilde{G} ({\bf n},{\bf n}^\prime ) \,\,
Y_{lm}({\bf n}^\prime )\,
d^2{\bf n}^\prime= \lambda_{\ell m}\,\, Y_{\ell m}({\bf n})\,\,,
\label{CP18}
\end{equation}
where 
\begin{equation}
\lambda_{\ell m} = \left\{ \begin{array}{ll}
(4\pi)^{-1} & \hspace{1cm} \textrm{if $(\ell, m) = (0,0)$} \\
(4\pi N)^{-1} & \hspace{1cm} \textrm{if $(\ell, m) \ne (0,0)$}
\end{array}
\right.\,.
\label{CP19}
\end{equation}
The eigenfunctions form a useful set for expansions of the intensity over the celestial sphere
\begin{equation}
I({\bf n})= \sum_{\ell = 0}^\infty \,\, \sum_{m = -\ell}^\ell\,\, a_{\ell m}\,  Y_{\ell m}({\bf n})\,\, .
\label{CP20}
\end{equation}
To incorporate the dependence on declination given in Eq.~(\ref{omeguita}),
let us re-define the angular intensity
\begin{equation}
I({\bf{n}}) = \frac{1}{\mathcal N}\,\,\sum_{j = 1}^N  \frac{1}{\omega_j} \,\, \delta ({\bf n}, {\bf n}_j) \,\,, 
\label{I}
\end{equation}
where $\omega_j$ is the relative exposure at arrival direction ${\bf n}_j$ and ${\mathcal N}$ 
is the sum of the weights $\omega_j^{-1}$. 
Since the eigenvalues of the $Y_{\ell m}$ expansion are 
uniquely defined
\begin{equation}
a_{\ell m} = \int  I ({\bf n}) \,\, Y_{\ell m} ({\bf n}) \,\, d^2{\bf n}\,\,,
\label{aintegral}
\end{equation}
the replacement of Eq.~(\ref{I}) into Eq.~(\ref{aintegral}) leads to the explicit form 
of the coefficients for our set of arrival directions 
\begin{equation}
a_{\ell m}=\frac{1}{{\mathcal N}}\sum_{j=1}^N \, \frac{1}{\omega_j} \,\, Y_{\ell m}({\bf n}_j) \,.
\label{CP21}
\end{equation}
With the coeffients given in this way, one can plot the intensity of the cosmic ray sky using Eq.~\ref{CP20}, 
as seen in Fig.~\ref{fig:intensity}. 
The mean square fluctuations of the coefficients are determined by the
power spectrum eigenvalues according to
\begin{equation}
\overline{a_{\ell m}^2}=\lambda_{\ell m}.
\label{CP22}
\end{equation}
 Although 
full anisotropy information is encoded into the coefficients $a_{\ell m}$ (tied to some 
specified coordinate system), the (coordinate independent) total power spectrum of fluctuations 
\begin{equation}
C(\ell) = \frac{1}{(2 \ell +1)}\,\, \sum_{m=-\ell}^\ell a_{\ell m}^2\,\,,
\label{CP23}
\end{equation}
provides a gross summary of the features present in the celestial distribution together with the 
characteristic angular scale(s). Note that Eqs.~(\ref{CP19}) and (\ref{CP22}) imply
\begin{equation}
\overline{C}(\ell) =  \frac{1}{(2 \ell +1)}\,\, \sum_{m=-\ell}^\ell \overline{a_{\ell m}^2}\,=
\left\{ \begin{array}{ll}
(4\pi)^{-1} &  \textrm{if $\ell = 0$} \\
(4\pi N)^{-1} & \textrm{if $\ell  \ne 0$}
\end{array}
\right.\,.
\label{Cp24}
\end{equation}
The power in mode $\ell$ is sensitive to variation over 
angular scales of  $\ell^{-1}$ radians.\cite{Sommers:2000us} Recalling that the 
estimated angular uncertainty for some of the events in 
the SUGAR sample is possibly as poor as $10^\circ$~\cite{Kewley:zt} we only look in this 
study for large scale patterns, going into the multipole expansion out to $\ell =5$.

\begin{figure}[tbhp]
  \begin{center}
	\leavevmode \epsfig{file=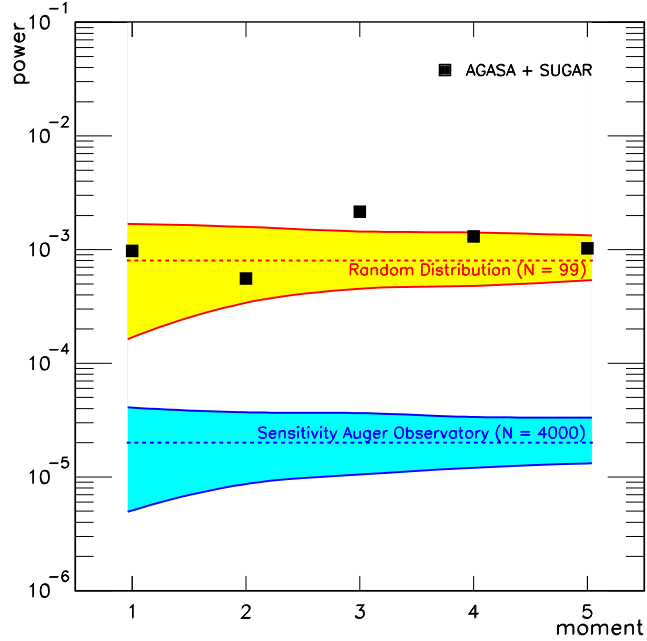,width=\regularwidthofpicture}
    \caption{
The angular power spectrum is indicated by the squares.  The horizontal lines
indicate the mean value, $\overline{C}(\ell) = (4 \pi N)^{-1},$ expected for an isotropic 
distribution. The upper shaded band shows the 1$\sigma$ fluctuation around the mean value for
$N=99$. The band was obtained from 1000 sets of Monte Carlo simulations
of 99 events each, including small corrections for $\omega_i$.  For $\ell=3$, where there is a 
small excess compared to the expectation for isotropy, $C(3) = 2.16\times 10^{-3}$ while 
the expectation from a random distribution  is $\overline{C}_{\rm MC} (3) = 9.5 \times 10^{-4}$, 
with a variance of $5.0 \times 10^{-4}$. The projected sensitivity for the Pierre Auger Observatory  
is also indicated on the plot by the lower shaded band.
      \label{fig4} }
    \end{center}
  \end{figure}

Our 
results at this juncture are summarized in Fig.~\ref{fig4}. The angular power spectrum is 
consistent with that expected from a random distribution for all (analyzed) multipoles, though there is a 
small ($2\sigma$) excess in the data for $\ell =3$. The majority of this excess comes from SUGAR 
data.\cite{Isola:2002ei} The decrease in error as $\ell$ increases may be understood as
a consequence of the fact that contributions to mode $\ell$
arise from variations over an angular scale $\ell^{-1}$.
If one compares to the expectation for isotropy, structures
characterized by a smaller angular scale, and hence
larger $\ell$, can be ruled out with
more significance than larger structures. 

To quantify the error, we study the fluctuations in $C(\ell)$ for $\ell \geq 1$. 
For simplicity, let us neglect the small effects of declination 
({\it viz.,} $\omega_i = 1 \,\,\forall i$), and consider the random variable
\begin{equation}
X_{\ell}=\frac{C(\ell)}{\overline{C}(\ell)}=
\left(\frac{4\pi N}{2\ell +1}\right)\sum_{m=-\ell}^\ell a_{\ell m}^2.
\label{A4}
\end{equation}
Denoting by \begin{math} P_\ell(\cos \delta)  \end{math} the Legendre polynomial
of order  \begin{math} \ell \end{math} and employing the addition theorem for
spherical harmonics,
\begin{equation}
\frac{4\pi }{2\ell+1}\sum_{m=-\ell}^\ell Y_{\ell m}({\bf n})Y_{\ell m}({\bf n}^\prime )
=P_\ell ({\bf n\cdot n}^\prime) \,,
\label{A5}
\end{equation}
Eqs.~(\ref{CP21}), (\ref{A4}), and (\ref{A5}) imply that
\begin{equation}
X_\ell=1+\frac{2}{N}\sum_{1\le i<j\le N}P_\ell({\bf n}_i\cdot {\bf n}_j).
\label{A6}
\end{equation}
Evidently, $\overline{X_\ell}=1$. Besides,
\begin{equation}
\overline{X_\ell^2}=1+\frac{4}{N^2}\sum_{1\le i<j\le N}\ \sum_{1\le k<q\le N}
\overline{P_\ell({\bf n}_i\cdot {\bf n}_j)P_\ell({\bf n}_k\cdot {\bf n}_q)}\,.
\label{A8}
\end{equation}
Since different pairs in the sum on the right hand side of Eq.~(\ref{A8})
are uncorrelated, it follows that
\begin{equation}
\overline{X_\ell^2}=1+\frac{4}{N^2}\sum_{1\le i<j\le N}
\overline{P_\ell({\bf n}_i\cdot {\bf n}_j)^2} \,.
\label{A9}
\end{equation}
There are \begin{math} \{N(N-1)/2\} \end{math} equivalent pairs in
Eq.~(\ref{A9}) which implies
\begin{equation}
\overline{X_\ell^2}= \overline{X_\ell}^2+2\left(1-\frac{1}{N}\right)
\overline{P_\ell({\bf n}_1\cdot {\bf n}_2)^2}\,.
\label{A10}
\end{equation}
From Eq.~(\ref{A5}) we obtain
\begin{eqnarray}
\overline{P_\ell({\bf n}_1\cdot {\bf n}_2)^2} & = &
\left(\frac{4\pi}{2\ell+1}\right)^2\sum_{m=-l}^l
\overline{Y_{\ell m}({\bf n}_1)^2}\ \overline{Y_{\ell m}({\bf n}_2)^2} \nonumber \\
 & = & \frac{1}{2\ell+1} \,.
\label{A11}
\end{eqnarray}
Plugging Eq.~(\ref{A11}) into Eq.~(\ref{A10}) leads to
\begin{equation}
\overline{\Delta X_\ell^2}=\left(1-\frac{1}{N}\right)\frac{2}{2\ell+1},
\label{A12}
\end{equation}
or equivalently, 
\begin{equation}
\left(\frac{\overline{\Delta C_\ell}^2}{\overline{C_\ell}^2}\right)
=\left(1-\frac{1}{N}\right)\frac{2}{2\ell+1} \,,
\label{A13}
\end{equation}
yielding (for large \begin{math} N \end{math})
\begin{equation}
\lim_{N\to \infty }
\left(\frac{\overline{\Delta C_\ell}^2}{\overline{C_\ell}^2}\right)
=\frac{2}{2\ell+1}\ \ {\rm for}\ \ \ell \ge 1\,,
\label{A14}
\end{equation}
which is the variance on $X_\ell$.

\section{Numerical Likelihood Analysis of Cosmic Ray Anisotropies}

The approach described above requires what might be described
as ``reasonably good statistics''. That is to say, one needs to 
have essentially {\em some} reasonable acceptance for every element of
solid angle in the sky. Problems of interpretation can easily
arise if there are regions which are unobservable -- blind spots --
and in this section I briefly describe how these can be dealt
with numerically. The work follows closely that presented in
\cite{ref:ICRC2003numerical_likelihood}.

\subsection{Spherical Harmonics}

It is convenient to define a set of real spherical harmonics
$\lbrace \psi_{\ell,m} \rbrace$ for $\ell=0,1,2,\ldots$ and 
$m=-\ell,-(\ell-1),\ldots,0,\dots,(\ell-1),\ell$
by
\begin{equation}
\psi_{\ell,m}(\theta,\phi) = \left\{ 
\begin{array}{ll}
k_\ell^{|m|}P_\ell^{|m|}(\cos\theta)\cos(m\phi) & \mbox{for\  $m=-\ell,\ldots,-1$},\\
k_\ell^0P_n^0(\cos\theta) & \mbox{for\  $m=0$,}\\
k_\ell^mP_\ell^{m}(\cos\theta)\sin(m\phi) & \mbox{for\  $m=1,\ldots,\ell$}
\end{array} \right. 
\end{equation}
\noindent
which are orthonormal with respect to the usual measure
$\sin(\theta)d\theta d\phi$ and integration over $\theta$ from 0 to $\pi$
and $\phi$ from 0 to $2\pi$, the $P_\ell^m$ are associated Legendre
polynomials of the first kind, and the normalization constants $k^\ell_m$ are:
\begin{equation}
k_0^0=\frac{1}{\sqrt{2\pi}}, k_\ell^0=\sqrt{\frac{2\ell+1}{4\pi}}, {\mathrm{\ and\ }}
k_\ell^m=\sqrt{\frac{2\ell+1}{2\pi}\frac{(\ell-m)!}{(\ell+m)!}}
\end{equation}
\noindent
The natural measure of anisotropy for a spherical 
distribution
is in terms of these spherical harmonics, as each $\ell$ labels, 
in a coordinate-independent fashion, just how much of each irreducible
$SO(3)$ representation is present.

In a perfect world with infinite statistics and complete sky coverage
there are now many possible approaches to estimating how much of each
of these components is present in a distribution, or, better,
what is the likelihood that a given function 
$f(\theta,\phi)$ with Fourier--Legendre expansion
$f(\theta,\phi)=\sum_{\ell=0}^\infty \sum_{m=-\ell}^{\ell}a_{\ell,m}\psi_{\ell,m}$ representing
the probability density of sources gives rise to the observed distribution
$g(\theta,\phi)$. The coefficients can be extracted from the usual
integral 
\begin{equation}
a_{\ell,m} = \int_0^{2\pi} \int_0^\pi f(\theta,\phi)\psi_{\ell,m}(\theta,\phi)\sin\theta d\theta d\phi
\end{equation}
\noindent
Of course this gives no measure of what sort of error should be associated with 
the determined values of each coefficient.
Alternative
approaches are to fit for the coefficients by minimizing some $\chi^2$-like
quantity like
\begin{equation}
\int_0^{2\pi} \int_0^\pi \frac{\left|g(\theta,\phi)- 
\sum_{\ell=0}^\infty \sum_{m=-\ell}^{\ell}a_{\ell,m}\psi_{\ell,m}(\theta,\phi)\right|^2}{\sigma^2}
\sin\theta d\theta d\phi
\end{equation}
\noindent
with $\sigma$ a suitable measure of error, 
or to compute and maximize a corresponding likelihood that the hypothesized
distribution parametrized by the $a_{\ell,m}$ gives rise to the the observed
distribution $g$.
Of course this is statistically unreasonable for a finite number of sources
({\em i.e.} to provide an infinite number of coefficients!), 
In general a decision must be made to truncate the expansion
at some value of $\ell$ to make the sum finite, but now the general phenomenon
of aliasing risks that, for example, a fit allowing for $\ell=0,1$ might
give misleading results for observed data which is drawn from a purely
$\ell=2$ distribution, say. 

A more serious problem is that should part of the sky be unobserved, there is now
no way to calculate anything! This is not a trivial point. An attempt to find 
anisotropies based only on observations in the Northern hemisphere with zeroes
inserted for the whole Southern hemisphere would be wildly
in error if some simple extrapolation were made to the unobserved region of the sky  --
especially if there were something bright and as-yet undetected in the South!
The real challenge is to say something statistically meaningful with the data
that is actually available. Clearly, observing only the Northern hemisphere and seeing
a good degree of isotropy should increase one's net belief in overall isotropy 
of the full sky, while leaving open the possibility of a staggeringly bright or
empty sky in the South.
One approach sometimes advocated is to try to make some sort of
new ``special functions'' which would be orthonormal over the
observed region of the sky, but it's not clear that this has much physical meaning as
it elevates a defect like ``lack of acceptance'' to a status comparable to ``the $SO(3)$ invariance
of space embodied in the $\psi_{\ell,m}$''. The following is a proposal for what
seems to make good statistical sense, and is physically unbiased.

\subsection{A Likelihood Proposal to Handle Limited Acceptance}

Based on the above observations, the following proposal seems reasonable: keep the spherical
harmonics as always with the (necessarily) truncated Fourier-Legendre expansion and construct an
unbinned likelihood\cite{Likelihoodbook} function in which one clearly specifies which values of $\ell$
are included in the sum. An unbinned likelihood function, as described below, automatically
makes maximum use of all detected information, allows for measurement errors to
be included easily, and is easy to implement numerically. Most importantly, however,
the likelihood is {\em not} to be normalized as it stands. Rather, one should
take the Bayesian approach
to likelihood which says that likelihoods give us ways to update our prior experimental
data or guesses in light of new information. This will mean that one can present
results on various hypotheses about data without bias, and with the easy inclusion
of other data from the same, or other experiments. 

To be concrete, we specialize here to the case where $g$ is a sum of
delta functions representing sources $i$ of unit intensity at $\theta_i,\phi_i$
and later discuss how one can treat the case of these being at uncertain
locations, or taking into account other properties such as intensity,
energy, composition, {\em etc.}. These will appear as natural generalizations
to the approach described.

The (unnormalized!) likelihood $L(\lbrace \ell\rbrace|\theta_i,\phi_i)$ that the measured $\theta_i,\phi_i$ arise
from $f(\theta,\phi)=\sum_{\ell}^\infty \sum_{m=-\ell}^{\ell}a_{\ell,m}\psi_{\ell,m}$
where the sum over $\ell$ is specified according to whatever hypothesis is being
tested ({\em i.e.} just taking $\lbrace\ell\rbrace=\lbrace 0,1\rbrace$ allows for uniform and dipole contributions
and no others, while $\lbrace\ell\rbrace=\lbrace2\rbrace$ would be pure quadrupole) is
\begin{equation}
L(\lbrace \ell\rbrace|\theta_i,\phi_i) = 
\frac{f(\theta_i,\phi_i)}{\int \prod_{{\ell}}\prod_{m=-\ell}^{\ell} da_{\ell,m} f(\theta,\phi) }
\end{equation}
\noindent
where the integral in the denominator is over all the parameters that can vary and over the
range in which the parameters are allowed to vary. An important caveat is that for $f$ to represent
a sensible probability distribution it should never be negative, and this must be checked.
Two approaches are possible: one is to restrict the domain over which coefficients range so
that the function is strictly positive (this is not actually very difficult in practice
since distributions are often nearly uniform with small fluctuations superimposed) or to 
take as a probability distribution function some positive function of $f$ in place of $f$
above. In astronomy~[4]
it is not uncommon to see $\exp(f)$. The choice ultimately represents the unavoidable
presence of some (often hidden) assumption about what a sensible prior ({\em i.e.}
in the absence of data) is for the likelihood -- a point to which we return later.

By construction then $L(\lbrace \ell\rbrace|\theta_i,\phi_i)$ is normalized so that its integral 
over all the parameters that can vary (the coefficients $a_{\ell,m}$ included in the truncated
Fourier-Legende expansion), and the total likelihood $L_{TOT}$, which is a function of those $a_{\ell,m}$ is
\begin{equation}
L_{TOT}(\lbrace \ell\rbrace)= \prod_i L(\lbrace \ell\rbrace|\theta_i,\phi_i)
\end{equation}
\noindent
This is a purely {\em relative likelihood}, 
and while no absolute normalization is possible (nor should it be!)
if part of the sky is unobserved, it is now very useful for two types of calculation. If one has
a prior expectation for the distribution of the $a_{\ell,m}$ (which might be that they are all
a priori equally likely) then $L_{TOT}$ can be multiplied by this and 
the product treated as a
normalized likelihood distribution
for the $a_{\ell,m}$ themselves. This is, of course, potentially dangerous, but does allow one
to see how the new data (the measured $(\theta_i,\phi_i)$) should cause one to revise earlier
beliefs. Such a likelihood can be maximized with respect to the $a_{\ell,m}$ and
if not enough data is present to test the hypothesis (for example, more 
coefficients
to fit for than data points) the fit will respond by just not converging 
(that is to
say, there will be flat directions in the likelihood as a function of the 
parameters
meaning that one can't decide) - the beauty of this approach is that it 
is, by
construction, correct.
When results are obtained, one automatically gets the values of the 
parameters, and  their entire likelihood distributions 
from which errors
(which need
not even  be Gaussian) can be extracted. One can even do things like fix some parameters 
to those given by a favoured theory and then repeat the fitting and 
then obtain likelihoods for the correctness of that theory.

More objectively, one can compute relative likelihoods in which the prior drops out,
so that it is reasonable to ask (even in the absence of full sky coverage!) what the relative
likelihood $L_{REL}$ is of pure dipole distribution to one admitting uniform, dipole and quadrupole
components:
%
$L_{REL} = \frac{L_{TOT}(\lbrace 1\rbrace)}{L_{TOT}(\lbrace 0,1,2\rbrace)}$
%
If one wants statements made to be about all energies over $x$, then one
just uses the
data points with energies over $x$. Similarly, data can be selected by
composition and (even relative) anisotropies be searched for as functions
of composition, energy, time, {\em etc.}
Extensions to uncertainty in direction are trivial to include: simply
divide a given event into
a large number, $M$, say of subevents  distributed appropriately and count 
each in the likelihood with
a weight $1/M$ -- this numerically convolves this uncertainty with the 
likelihood, and the size of any additional errors introduced by the procedure
can be studied by varying $M$.

\section{Irresponsible Speculations}

It is interesting to speculate as to what any possible anisotropies
might mean, so focussing on the idea of a few bright sources turning up,
let us indulge in one wild idea, more for amusement
than anything else! 

History has seen many attempts by human beings to either detect messages from
extraterrestrial civilizations, or to send messages that might be received
by intelligent life elsewhere. Carl Sagan, Linda Salzman Sagan, and Frank Drake\cite{Pioneer}
had already proposed in 1972 that Pioneer 10, the first manmade craft to leave our
solar system, should carry a message-bearing plaque in case some alien stumbled
across it -- and it did! This was a big step up from earlier proposals
including one attributed to Carl Friedrich Gauss in the 1820's which suggested
laying out geometrical patterns in the vast forests of Siberia by cutting our
huge swaths of trees and planting wheat in their place. Austrian astronomer Johann
Joseph von Littrow is alleged to have suggested the similarly non-environmentally
friendly alternative of digging a vast circular canal (miles across) into the Sahara Desert,
filling it with kerosene, and setting it on fire in the hopes that someone or
something would notice and be impressed.

Suggestions of radio-based communication go back to Nikola Tesla in the late
1800's, and progressed through Frank Drake's 1960 ``Project Ozma'' to listen for
signals at the 21 cm line of hydrogen to the modern SETI program. The interested
reader can find a lot of historical data at the website \cite{SETI}.

Personally, I think it's interesting to wonder what would constitute an
``impressive'' signal to, or from, another civilization. By way of
comparison, I am quite aware of
the fact that gorillas have some degree of linguistic ability, but I am quite
unlikely to spend much time trying having a sophisticated conversation with
a gorilla. While a gorilla and his friends may be quite impressed with themselves,
the simple fact is that I'm unlikely to take much notice of their ``intellectual''
achievements, and will think of them simply as apes. (There is, of course,
no animosity intended -- I'm just saying...)

On the other hand, if a gorilla were to construct a laser pointer and flash
that around a bit, I might be a lot more impressed -- at least here's a gorilla
that has managed to make a controlled source of coherent light! In fact, the optical
version of SETI\cite{opticalSETI} makes the move away from looking for (or sending)
signals by radio to using light. This, of course, offers the tremendous advantage that
one can direct a laser towards a potentially inhabited planet and use a tiny
fraction of the energy that would be required for a non-coherent source. Presumably
{\em that} would be impressive to an alien civilization, and then they'd {\em really}
want to talk to us -- or would they?

It's easy to argue that more exotic physics would be required for us to 
reach the threshold of being ``interesting''. Perhaps we are awash in 
low energy neutrino signals or gravitational waves for which we have no suitable detection 
devices. As far as we know, these {\em might}
be possible channels for communication given a suitable trick of
engineering -- maybe we just haven't come up with it yet. It's certainly
possible that the technologies needed to produce and detect these signals
would develop hand-in-hand, in which case these channels would only
be useful between civilizations at comparable levels of technology,
and we would be able to participate in interstellar communications about
the same time as we were able to listen in.

UHECR's present a rather different kind of possible message-carrier:
within our current knowledge of physics there's no way to make
these with a device that would fit on our planet, while on the other hand,
we are quite able to detect them! 

If it turns out that there
is no natural mechanism of any kind to make UHECR's, and they {\em must}
be created by artificial devices of a rather advanced technological nature,
they would make great conversation openers from aliens who want
to get our attention, even though we may not have comparable technology 
to signal back. Perhaps an anisotropy which
revealed itself as pointing back to a planetary system could be a message
to the effect of ``Hey! Look at me! I've got advanced technology -- advanced
enough to make protons at $10^{20}$ eV and get them to you! These are pretty
easy for you to detect (even though they're hard for me to produce), but at least you
know I'm here!'' In this sense, one might imagine that intelligently generated 
UHCER's could be cosmic signal flares to get our attention!

This isn't really a completely serious suggestion, but then it's not entirely
{\em not} serious. The only time it seems to have been discussed
semi-seriously is in bars in Rio after this talk!

\section{Concluding Remarks}

I have reported on the first full-sky anisotropy search using
data from the SUGAR and AGASA experiments.  At present, low statistics
and poor angular resolution limits our ability to perform a very sensitive
survey, but we can at least have a preliminary look at the first moments in the
angular power spectrum. The data are consistent with isotropy, though
there appears to be a small excess for $\ell=3$, arising mostly from the
SUGAR data.

There are two caveats in this analysis which should be kept in mind.
First, from the published SUGAR results, it is difficult to make an
exact determination of the exposure, as the sensitive area of the experiment
varied as a function of time.  Here, we assumed an area-time product of
approximately 775~km$^2$ yr.  Second, there is some uncertainty in
the energy calibration.  The SUGAR results are reported in terms of the
number of vertical equivalent muons together with two possible models to
convert this to primary energy.  We have chosen the model yielding an energy
spectrum which is in better agreement with the AGASA results~\cite{Anchordoqui:2003gm}.  
It should be noted, though, that this spectrum does not agree well with the results
of the High Resolution Fly's Eye (HiRes) experiment.\cite{Abu-Zayyad:2002ta} 
Though there are uncertainties in 
the energy scale, the impact on this anisotropy search may not be so severe.  This is because
the energy cut of $10^{19.6}$~eV is well above the last break in the spectral
index at $\sim 10^{18.5} - 10^{19}$~eV, and one would expect that all cosmic rays
above this break share similar origins.

\begin{figure}[thpb]
\begin{center}
\includegraphics[width=0.7\textwidth]{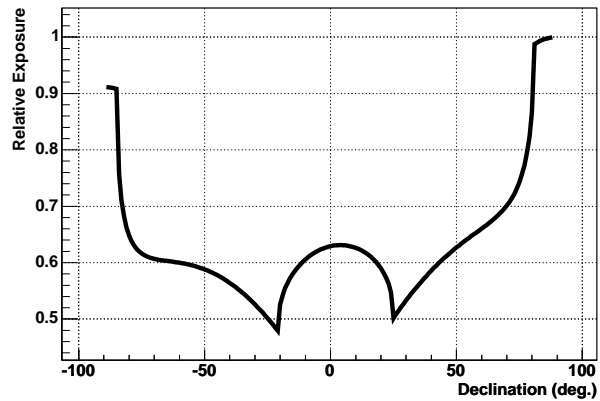}
\caption{Relative exposure of the full Pierre Auger Observatory.}
\label{figure:paoexp}
\end{center}
\end{figure}

In the near future we expect dramatically superior results from
the Pierre Auger Observatory\cite{AUGERwww}. This observatory is designed to measure the energy and arrival direction 
of ultra high energy cosmic rays with unprecedented precision. It will consist 
of two sites, one in the Northern hemisphere and one in the Southern, each covering an area 
$S \approx 3000$~km$^2$. The Southern site is currently
under construction while the Northern site is pending. Preliminary results from the Southern site's
engineering array are available in reference \cite{AUGER}.
Once complete, these two sites together
will provide the full sky coverage and well matched exposures
which are crucial for anisotropy analyses. Using Eq.~\ref{omeguita}, the calculated relative exposure is seen in Fig.~\ref{figure:paoexp}. 
The base-line design of the detector includes a ground array 
consisting of 
1600 water \v{C}erenkov detectors overlooked by 4 fluorescence eyes. The angular and energy resolutions 
of the ground arrays are typically less than $1.5^\circ$ (multi-pole expansion $\ell \sim 60$) 
and less than 20\%, respectively. The detectors are designed to be fully 
efficient ($p \approx 1$) out to $\theta_{\rm max} = 60^\circ$ beyond $10^{19}$~eV, yielding a 
nearly uniform sky $A \approx 1.4 \times 10^4$~km$^2$~sr.\cite{Sommers:2000us} In 10~yr of running the 
two  arrays will collect $\approx 4000$ events 
above $E_1 = 10^{19.6}$~eV. As can be seen from Fig.~\ref{fig4}, such statistics will allow one to 
discern asymmetries at the level of about 1 in $10^4$.

\section*{Acknowledgments}

It is a great pleasure to thank my collaborators, in particular Luis Anchordoqui
who contributed enormously to the preparation of this writeup, and
Carlos Hojvat, Tom McCauley, Tom Paul, Steve Reucroft, and Allan Widom with whom
much of this work was done, as well as all my colleagues in the Pierre Auger
Collaboration, especially Tere Dova and Paul Sommers. 
I would also like to thank the conference
organizers, especially Santiago Perez Bergliaffa, Mario Novello, and Remo Ruffini
for their hospitality, and everyone in Rio, the most beautiful city in the world!
Muito obrigado!


\begin{thebibliography}{99}
\bibitem{ylm} For this analysis 
we use real-valued spherical harmonics, which are obtained from the complex ones by substituting, 
$e^{i\,m\,\phi} \rightarrow \sqrt{2} \sin (m\phi)$, if $m<0$, $e^{i\,m\,\phi} \rightarrow \sqrt{2} 
\cos (m\phi)$, if $m>0$, and $e^{i\,m\,\phi} \rightarrow 1$ if $m=0.$ These
are discussed again in more detail in section 4.1 where the fact that they
are real is of paramount importance in constructing a real--valued
likelihood function.
\end{thebibliography}
\end{document}